\documentclass[lettersize,journal]{IEEEtran}
\usepackage{amsmath,amsfonts}
\usepackage{amssymb}
\usepackage{algorithmic}
\usepackage{algorithm}
\usepackage{array}
\usepackage[caption=false,font=normalsize,labelfont=sf,textfont=sf]{subfig}
\usepackage{textcomp}
\usepackage{stfloats}
\usepackage{url}
\usepackage{verbatim}
\usepackage{graphicx}
\usepackage{cite}
\usepackage{booktabs}
\usepackage{xcolor}
\usepackage{multicol}
\usepackage{multirow}
\begin{document}

\title{Hierarchical Multicriteria Shortest Path Search}

\author{Temirlan Kurbanov
\thanks{The authors are with the Department of Computer Science, Faculty of Electrical Engineering, Czech Technical University in Prague (email: kurbatem@fel.cvut.cz).

This work was co-funded by the European Union under the project ROBOPROX - Robotics and Advanced Industrial Production (reg. no. CZ.02.01.01/00/22\_008/0004590).}, Linxiao Miao, Ji\v{r}\'{i} Vok\v{r}\'{i}nek}

\maketitle

\begin{abstract}
This paper presents a novel multicriteria shortest path search algorithm called Hierarchical MLS. The distinguishing feature of the algorithm is the multilayered structure of compressed k-Path-Cover graphs it operates on. In addition to providing significant improvements in terms of time and memory consumption, the algorithm is notable for several other features. Due to the preprocessing phase requiring only several seconds, the algorithm can be successfully applied to scenarios with dynamic prices. Moreover, the algorithm does not employ bidirectional search, and can thus work on time-dependent metrics. We test the algorithm on multiple graphs and analyze its performance in terms of time and memory efficiency. The results prove Hierarchical MLS to be faster than its direct alternatives by at least 2 times in terms of query runtime and at least 20 times in terms of preprocessing.
\end{abstract}


\section{Introduction}
\IEEEPARstart{M}{ulticriteria} optimization is a discipline concerned with finding solutions optimizing multiple objective metrics at once. Applied to shortest path search, the problem aims to find routes that satisfy several criteria (e.g. time, distance, fuel consumption) as well as possible. Naturally, the criteria in question are often in conflict with each other. Thus, the result of multicriteria optimization is a Pareto set of viable solutions, each of which represents a possible trade-off between the criteria. Although the problem is NP-hard \cite{complexity} and thus very hard to effectively apply in practice, it boasts inherent advantages as opposed to the simpler approaches of constrained optimization and linear combination of criteria. Namely, multicriteria optimization provides the full spectrum of possible solutions, giving the user the ability to choose one based not only on their preference, but also on the relative quality of the solution set.

With the increasing demand on sustainable cities and various green initiatives, multicriteria shortest paths search has been receiving increasing attention. Despite a substantial body of research, each of the proposed improvements has its limitations. For instance, some of the methods rely on simultaneous bidirectional search from the origin and goal positions. Unfortunately, these methods cannot be applied when one of the metrics is time dependent. For instance, a backward search from the goal based on traffic-dependent traversal time is impossible, since one cannot know when the goal will be reached and what the traffic conditions will be in this time instance. Other methods make use of various preprocessing approaches: graph preprocessing, limit precomputation, and so on. The issue is, however, that these procedures are often expensive and therefore cannot be performed often, making them inapplicable to scenarios with dynamic metrics.

In this paper, we propose an algorithm we call Hierarchical MLS. The core of this approach is in a modified MLS \cite{Martins} shortest path search that operates on the multilayered structure of Hierarchical k-Path-Covers (kPCs) \cite{HierKPC}. Although the building procedure for this structure is a preprocessing step, the smart design of it allows the procedure to take only several seconds on country-sized road graphs. As such, the proposed algorithm is free of the drawbacks described earlier. Furthermore, it provides significant efficiency gains both in terms of time and memory requirements and can further incorporate the majority of existing speedup techniques.

To demonstrate the claimed superior performance of the algorithm, we test it on multiple road network instances, including country-sized graphs of Germany and road instances from the 9th DIMACS challenge.

The remainder of the paper is organized as follows. Section 2 considers related work in the are of multicriteria shortest path search. Section 3 is dedicated to problem formulation and methodology description. Section 4 describes the experimental analysis and its results, while Section 5 gives a conclusion for the paper. 

\section{Related Work}
In recent years, a substantial amount of research effort has been dedicated to the multicriteria shortest path search problem. Nevertheless, all algorithms able to solve it to optimality can be traced back in their principles to Multicriteria Label-Setting (MLS) \cite{Martins} and Multicriteria Label-Correcting (MLC) \cite{Vincke} algorithms. Even these two algorithms are similar in all aspects but one: MLS enforces lexicographically ordered processing of solutions, or "labels", while MLC processes them in FIFO fashion. This allows MLS to only process each label once and "set" it, while MLC often goes back to an already processed label and "corrects" it.

Despite being based on the same core principles, some of the newly proposed algorithms introduce significant modifications to them. BBDijkstra \cite{BBDijkstra}, for instance, is a bidirectional MLS search combined with several pruning heuristics, providing a substantial speedup over the classic approaches. In \cite{parallel}, on the other hand, the inner label expansion procedure was parallelized. Interestingly, the proposed parallelization method could be combined with most other label-setting algorithms. MDA \cite{MDA, MDA2} is another improvement of MLS employing intelligent label expansion and maintenance techniques, which allows it to have a smaller memory footprint and decreased operation time. $t$-discarding kPC-MLS \cite{t-kPC-MLS} is a combination of MLS with two other methods. It uses a kPC building algorithm developed by Funke et. al. \cite{kPC, kPC_pruning} to construct a single-layer k-Path-Cover. The operation of MLS on this cover is improved by a dimensionality reduction technique \cite{dimensionality}. The resulting combination is particularly suited for operation on larger road networks, where it outperforms other alternatives. Other improvement approaches for multicriteria shortest path search also include guiding heuristics like the ones employed in A$^*$. This is done, for example, in MOA$^*$ \cite{MOA*} and NAMOA$^*$ \cite{NAMOA*}.

Another direction for the research on the problem is heuristic and metaheuristic methods. Naturally, these trade solution optimality for faster operation, and are thus preferential on larger problem instances. For instance, several pruning heuristics have been researched in \cite{heur_t-kPC-MLS}. Among examples of metaheuristic approaches are simulated annealing \cite{SA} , ant colony \cite{ant}, and evolutionary algorithms \cite{bi_ev}.

\section{Methodology}
This section gives a formal description of the multicriteria shortest path search problem and presents the details of the proposed algorithm.
\subsection{Problem formulation}
Let $G = (V, E)$ be a finite directed graph consisting of $|V|$ vertices and $|E|$ edges. Every edge $e = (v_i, v_j)$ has a starting vertex $v_i$, an ending vertex $v_j$, and a tuple $\gamma(e)$ of cost values representing the criteria we want to optimize. A path $\pi$ can be represented either by its compound vertices $\pi = v_1, \dots, v_n$ or by its compound edges $\pi = e_1, ..., e_{n-1}, \forall i \in \{1, n-1\}: e_i = (v_i, v_{i+1})$. The cost tuple of a path is calculated from the tuples of its compound edges. Probably the most ubiquitous for this purpose is the sum-type criterion function $\gamma^{\pi} = (\sum_e \gamma_1^e, \dots, \sum_e \gamma_q^\pi), e \in \pi$, which we will also be using in this paper. The task of multicriteria shortest path search is, for given origin and destination locations, to find the Pareto set of paths between them. 

The Pareto set is defined using the dominance property. For two tuples $\gamma = ( \gamma_1, ..., \gamma_q ) \in \mathbb{R}^q$ and $\gamma'=( \gamma_1', ..., \gamma_q') \in \mathbb{R}^q$, the relation of \textit{weak dominance} of $\gamma'$ by $\gamma$ is defined as follows:
\begin{equation}
	\begin{gathered}
		\gamma \preceq \gamma' \textrm{ iff:} \\
		\forall i \in \{ 1, ..., q \}
		\begin{cases}
			\gamma_i \leq \gamma_i' & \textrm{ if criterion } i \textrm{ is minimized}\\
			\gamma_i \geq \gamma_i' & \textrm{ if criterion } i \textrm{ is maximized}.
		\end{cases}
	\end{gathered}
\end{equation}
Additionally, $\gamma$ \textit{dominates} $\gamma'$ ($\gamma \prec \gamma'$) iff $\gamma \preceq \gamma'$ and $\gamma' \npreceq \gamma$. Since there is no need for us to distinguish two paths with exactly equal costs, we use weak dominance for our computations.
Having established the relation of weak dominance, we can now say that a Pareto set consists of individual solutions (cost tuples in this case) that are not (weakly) dominated by any other solution. Thus, a Pareto set provides a selection of trade-offs between the considered criteria such that none is better than any other one in all aspects.

\subsection{Hierarchical Multicriteria Label-Setting Algorithm}
As one can see from Section II, there are several algorithms able to solve the multicriteria shortest path search problem, the most basic ones being MLS \cite{Martins} and MLC \cite{Vincke}. In a way, these algorithms can be considered extensions of Dijkstra's algorithm to the multicriteria setting, where they operate not on distances, but labels, each of which represents a path from the origin to the vertex the label is assigned to. More recent notable solutions include MDA \cite{MDA}, a more lightweight version of MLS, and $t$-discarding kPC-MLS \cite{t-kPC-MLS}, a combination of methods designed for one-to-many multicriteria shortest path search on large road networks.

Here, we propose the Hierarchical Multicriteria Label-Setting Algorithm (Hierarchical MLS). In its main aspects, it is similar to $t$-discarding kPC-MLS: it has a preprocessing phase based on k-Path-Covers and uses a modified MLS with a dimensionality reduction technique \cite{dimensionality}. The key distinction of this algorithm, however, is the nested structure of hierarchical $k$-Path Covers it operates on. This structure is built using a different approach than the one in t-kPC-MLS, and MLS is modified to operate on the multilayer structure as a whole, not on a single level of it. 

For a graph $G = (V, E)$, its $k$-Path Cover (kPC) $V^k \subseteq V$ is a subset of its vertices such that \textit{for every simple path of size $k$ in $G$, at least one of its vertices must be in the cover}. From this cover, we create a cover graph $G^k = (V^k, E^k)$, where $E^k$ is the set of cover edges connecting $V^k$ and representing non-dominated paths between these vertices in $G$. 
The process of building hierarchical covers largely follows the one explained in \cite{HierKPC}, except extended to the multicriteria setting.

The idea of building hierarchical kPCs lies in building nested covers, for instance $G^2$ from $G$, $G^4$ from $G^2$, $G^8$ from $G^4$, and so on. Theoretically, any value of $k$ can be used for this nested structure, but setting it to 2 allows us to use much simpler procedures and thus speed up the whole building process. From now on, we will denote the hierarchical cover graphs using subscripts $G_p$, where $k = 2^p$. Thus, $G=G^1=G_0$, $G^2 = G_1$, and so on. Assuming we build $\mathcal{T}$ levels, the top level will be $G_\mathcal{T}$. In \cite{HierKPC}, several strategies for 2-path vertex cover building are described and compared. We, however, used the strategy the authors call LR-deg due to it consistently creating smaller-sized covers than the alternatives. For LR-deg, the cover vertex set is initialized to an empty set. The vertices are processed in the increasing order of their degrees, and for every vertex that is not in the cover, the set of its neighbors is added to it.

The procedure for cover edge building is also significantly simplified for the 2-sized covers. In essence, it suffices to go through every vertex that was not added to the cover and examine every pair of its incoming and outgoing edges. For every such pair, if it is not dominated by an existing cover edge between the start and end vertex, it can be added to the cover. In case of city-route planning, it is also necessary to make sure that a vehicle can progress from the incoming edge to the outgoing one.

The methods described above can be used on a $G_t$ cover to build $G_{t+1}$. Thus, a hierarchical cover structure can be built from a graph $G$ by consecutively applying the cover-building procedures to the top layer until a desired level is reached. Since these procedures are relatively simple and can operate very efficiently, hierarchically building a high-level cover is faster than building one from scratch. This is demonstrated by the experiment results in \cite{HierKPC}.

The nested kPC structure has the original graph at its base level as $G_0 = G$. This allows the algorithm to contain the full information on the graph and start from any location in it. The core idea of the proposed method is that an adapted MLS query starts from the start and goal positions in the nested structure and climbs to the highest kPC level as fast as possible, where the bulk of the searching is performed.

MLS is a graph search algorithm operating on \textit{labels}. Each label represents a path from a source point to the vertex the label is associated with. As such, the label contains the information on the route criteria of the path it represents. Every vertex $v \in V$ in MLS has two sets of labels associated with it: permanent $perm(v)$ and temporary $temp(v)$. The algorithm starts with a priority queue $Q$ of lexicographically ordered labels. Given two vectors $\gamma = (\gamma_1, ..., \gamma_q)$ and $\gamma' = (\gamma_1', ..., \gamma_q')$, we say that $\gamma$ \textit{lexicographically precedes} $\gamma'$ iff:
\begin{equation}
	\begin{gathered}
		\exists j \in \{1, ..., q\}:
		\{\forall i \in \{1, ..., j-1\}: \; \gamma_i = \gamma_i'\} \textrm{ and} \\
		\begin{cases}
			\gamma_j < \gamma_j' & \textrm{if } j \textrm{ is minimized}\\
			\gamma_j > \gamma_j' & \textrm{if } j \textrm{ is maximized}\\
		\end{cases}
		\\
		\textrm{OR}
		\\
		\forall i \in \{1, ..., q-1\}: \gamma_i = \gamma_i' \; \textrm{ and} \\
		\begin{cases}
			\gamma_q \leq \gamma_q' & \textrm{if } q \textrm{ is minimized}\\
			\gamma_q \geq \gamma_q' & \textrm{if } q \textrm{ is maximized.}\\
			
		\end{cases}
	\end{gathered}
\end{equation}
The source label is created and put into the temporary set of the source vertex and into the priority queue. During every query iteration, the lexicographically smallest label in $Q$ is removed from it, moved from the temporary set of its vertex to the permanent one, and expanded to neighbor vertices. The newly generated labels are then put into their respective temporary sets and the priority queue if they are not dominated by the temporary and permanent sets of their vertices. This loop iterates until the priority queue is empty or a termination condition is hit. The lexicographic ordering of the labels makes the algorithm label-setting as opposed to label-correcting and guarantees that permanent sets only contain labels that belong to the Pareto set.

The hierarchical kPC structure means that a single vertex can appear on multiple levels. In fact, every vertex $v \in V$ has a top level $T(v)$ such that $v \in V_t, \forall t \leq T(v)$ and $v \notin V_t, \forall t > T(v)$. A key feature of Hierarchical MLS is that label expansion from a vertex $v$ is always conducted on level $T(v)$, which allows the algorithm to quickly advance to the top level of the nested covers.  Given a query of start and destination vertices $(s, d)$, Hierarchical MLS operates in two stages. The goal of the first stage is to connect the destination vertex to the top cover level. This is done by using backward expanding Hierarchical MLS. The algorithm starts from the goal vertex and proceeds backwards through connected edges. Besides operating in reverse direction, the query proceeds normally until all its labels are expanded to $G_\mathcal{T}$. The first stage of the algorithm progresses until the priority queue is empty and gives as a result sets of labels for "hit" vertices in the topmost cover leading to the goal vertex. Naturally, the first stage is skipped if the goal vertex belongs to the top cover. If the criteria values for the search from the destination vertex are known in advance, the first stage can include standard domination checks to remove dominated labels. If, however, the algorithm operates with criteria that cannot be calculated prior to reaching the edge in question (for instance, traffic density on the edge will depend on the exact time it is traversed), then it suffices to just expand all labels without domination checks. The partial solutions belonging to the Pareto set will, in this case, be determined when the second stage reaches the hit vertices. 

The second stage of Hierarchical MLS starts from the source vertex and expands along the direction of the edges. As it was stated earlier, every expansion from a vertex $v$ is done on its top level $T(v)$. Whenever MLS expands a label from a vertex $v$ to its neighbors, it first determines the top level $T(v)$ the vertex is present in. The expansion is thus performed only on this level. Applying this principle to 2-Path-Covers means that every expanded label is at least one level higher than its predecessor. In practice, however, it is not unusual for a label from a vertex $v, T(v) = 0$ to be expanded to a neighbor vertex $w, T(w) = \mathcal{T}.$ When the query reaches the top level, further expansion is only performed on it. Whenever a label is expanded to a hit vertex, it is automatically extended to the destination vertex through the computed backward paths. These full paths are added in a separate set of labels where only non-dominated solutions are kept. As usual, the Hierarchical MLS progresses until the priority queue is empty or the termination condition is hit.

Additionally, we augment our Hierarchical MLS with dimensionality reduction via $t$-discarding \cite{dimensionality}, which is researched in more detail in \cite{t-kPC-MLS}. It uses $t$sets --- structures containing non-dominated truncated vectors $\gamma^t$ of the permanent labels of the edge. For a criteria vector $\gamma = (\gamma_1, ..., \gamma_q)$, its truncated version is $\gamma^t = (\gamma_2, ..., \gamma_q)$. $t$sets are used in the $t$-discarding procedure, which decreases the number of operations necessary for dominance checks. This is assured by two factors. Firstly, due to the labels being expanded lexicographically, comparison of the first criterion (in our case travel time) is unnecessary, since permanent labels were created earlier than the newly expanded one. Secondly, due to the decreased number of criteria, the set of non-dominated truncated vectors is usually significantly smaller than the original. In \cite{dimensionality}, it is theoretically proven that $t$-discarding can replace classic dominance checks against the set of permanent labels provided the algorithm operates in the lexicographic order. The pseudocode of the algorithm is available in Algorithm \ref{alg:multicriteria_path}.
\begin{algorithm}[tbh]
    \caption{\textit{HierarchicalMLS($\mathcal{H}, s, d$)}}
    \label{alg:multicriteria_path}
    \textbf{Input}: hierarchical covers $\mathcal{H} = (G_0, ..., G_\mathcal{T})$, source vertex $s \in V$, destination vertex $d \in V$\\
    \textbf{Output}: Pareto set of routes from source $s$ to destination $d$
    \textbf{Parameters}: local area $A$
    \begin{algorithmic}[1] 
        \STATE $perm(v) = \emptyset \quad \forall v \in V$
        \STATE $temp(v) = \emptyset \quad \forall v \in V$
        \STATE $tset(v) = \emptyset \quad \forall v \in V$
        \STATE $compositeRoutes = \emptyset$
        \STATE $backwardRoutes = backwardMLS(\mathcal{H}, d)$
        \STATE $Q =$ empty lexicographic queue
        \STATE $Q.enqueue(\text{source label})$
        \STATE $temp(v).add(\text{source label})$
        \WHILE{$Q$ is not empty}
            \STATE $l = Q.dequeue()$
            \STATE $v = l.vertex$
            \STATE $t = T(v)$
            \STATE $temp(v).remove(l)$
            \STATE $perm(v).add(l)$
            \STATE $tset(v).update(l)$
            \IF{$v \in backwardRoutes$}
                \STATE $possibleRoutes = $
                \STATE \qquad $connectRoutes(l, backwardRoutes(v))$
                \STATE $compositeRoute.update(possibleRoutes)$
            \ENDIF
            \FOR{$e \in G_{t}.\delta^+(v)$}
                \STATE $v' = e.destination$
                \STATE $l_{new} = expand(l, e)$
                \IF{$tset(v') \text{ $t$-discards } l_{new}$}
                    \STATE \textbf{continue}
                \ENDIF
                \IF{$\exists l' \in temp(v'): l' \preceq l_{new}$}
                    \STATE \textbf{continue}
                \ENDIF
                \STATE $toRemove = \{l' \in temp(v'): l_{new} \preceq l'\}$
                \STATE $temp(v') = temp(v') \setminus toRemove$
                \STATE $temp(v').add(l_{new})$
                \STATE $Q.enqueue(l_{new})$
            \ENDFOR
        \ENDWHILE
        \STATE \textbf{return } $compositeRoutes$
    \end{algorithmic}
\end{algorithm}

Another important feature of the proposed algorithm is its ability to adopt the majority of existing pruning and heuristic techniques. A notable method that we used in our experiments is limit precomputation \cite{bounds}. The idea of the method is to use backward Dijkstra queries for every criterion and thus build upper and lower bounds to the destination for every vertex in the graph. Then, during the multicriteria query, a candidate label is extended with the lower bound from the vertex to the destination. The resulting estimate is domination-checked against the upper bounds of the source vertex, and the label is discarded if the estimate is dominated by the source bounds. In case of Hierarchical MLS, bound computation and usage is modified to work on the nested kPCs. For single-criterion Dijkstra searches, the origin vertex is first expanded from to compute its "hit" labels in the top level. In the second stage, a backward Dijkstra query is performed from the destination using the same upward principle of expanding only on the top level. Since the bulk of the multicriteria query is performed on the top level $G_\mathcal{T}$, the bounds are saved and used only for the vertices in it and the origin vertex. 

\section{Experiments}
This section gives a description on conducted experiments and their results. The aim of the experiments was to measure the size and construction time of hierarchical k-Path-Covers and analyze the running time and memory requirements of Hierarchical MLS.

\subsection{Experiment Setting}
We tested the proposed algorithm on two groups of graphs. The first group is taken from OpenStreetMap and represents road networks of Bavaria\protect\footnotemark \footnotetext{https://download.geofabrik.de/europe/germany/bayern.html} (BAV) and Germany\protect\footnotemark \footnotetext{https://download.geofabrik.de/europe/germany.html} (GER). Residential roads were removed from the graphs, and the criteria we tested the algorithm on were traversal time and energy consumption for electric vehicles computed using simple estimation functions. The second group of graphs belongs to the $9^{\text{th}}$ DIMACS implementation challenge\protect\footnotemark \footnotetext{http://users.diag.uniroma1.it/challenge9/download.shtml}. The graphs represent multiple areas of the USA: New York City (NY), San Francisco Bay (BAY), Colorado (COL), Florida (FLA), and Northeast USA (NE). The planning criteria were distance and traversal time. The sizes of the graphs are listed in Table \ref{tab:sizes}. The algorithms were implemented in C++20, using only standard libraries of the language. The tests were conducted on a server using AMD EPYC 7543 at 3.1 GHz, and no parallelization was used.
\begin{table}[t]
	\caption{graph sizes}
	\begin{center}
		\label{tab:sizes}
		\begin{tabular}{|c|r|r|}
			\hline
			graph & vertex \# & edge \#\\ 
			\hline
			BAV & 294727 & 587782 \\
			GER & 1521776 & 2947009 \\
            NY & 264346 & 733846 \\
			BAY & 321270 & 800172 \\
            COL & 435666 & 1057066 \\
            FLA & 1070376 & 2712798 \\
            NE & 1524453 & 3897636 \\
			\hline
		\end{tabular}
	\end{center}
\end{table}

\subsection{kPC experiments}
\begin{figure}
	\centering
    \includegraphics[width=\linewidth]{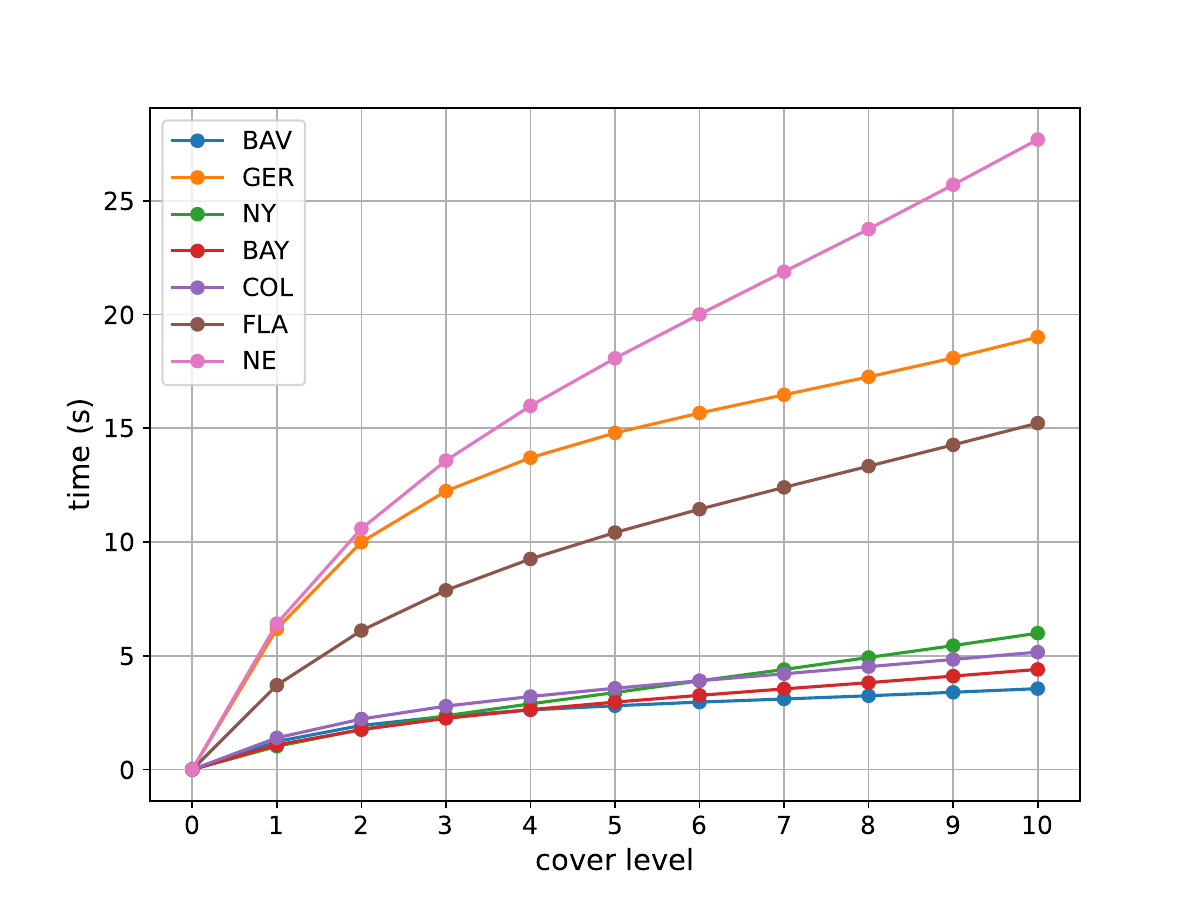}
    \caption{Cumulative construction times of hierarchical covers}
    \label{fig:kpc_times}
\end{figure}

\begin{figure}
    \centering
    \includegraphics[width=\linewidth]{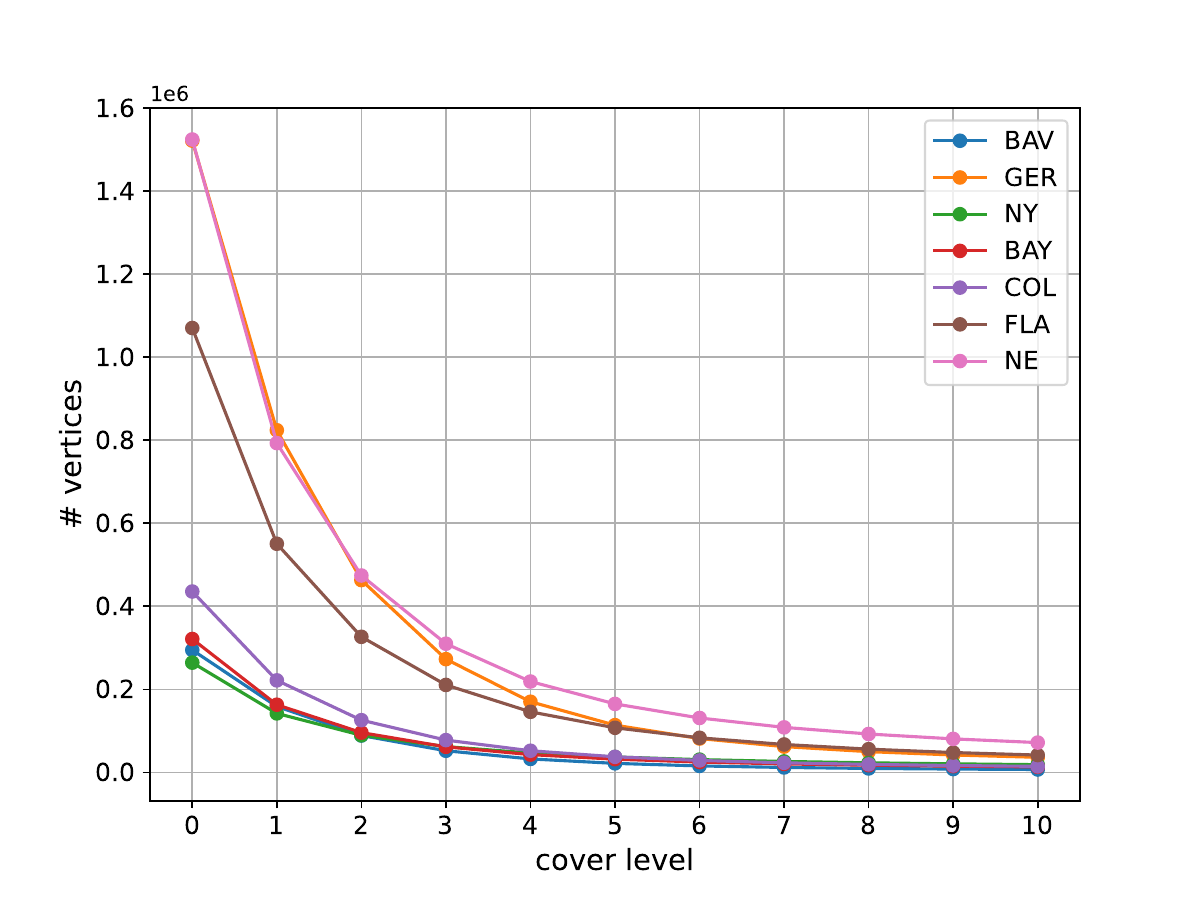}
    \caption{Number of vertices of hierarchical covers}
    \label{fig:kpc_vertices}
\end{figure}

\begin{figure}
    \centering
    \includegraphics[width=\linewidth]{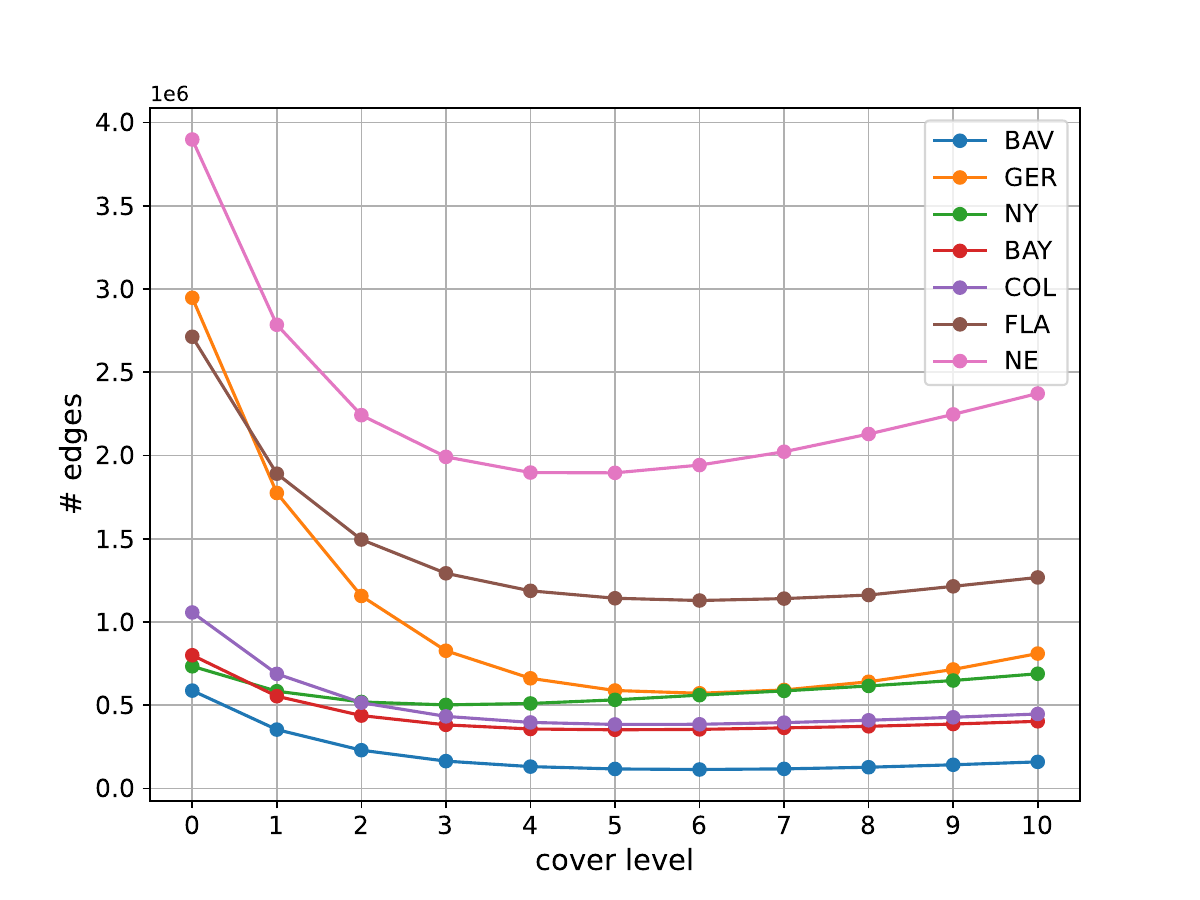}
    \caption{Number of edges of hierarchical covers}
	\label{fig:kpc_edges}
\end{figure}

For all of the tested graphs, we built 10 levels of hierarchical covers beside the original graph set at position 0. We measure the construction times and sizes of the covers, with the results being presented in Figures \ref{fig:kpc_times}, \ref{fig:kpc_vertices}, and \ref{fig:kpc_edges}.


\begin{table}[t]
	\caption{Comparison of single-layer and hierarchical $k$PC. $k$ is set to 32 for BAV and GER, 16 for NY, BAY, and COL.}
    \label{tab:kpc_comp}
	\begin{center}
		\label{sizes}
		\begin{tabular}{|c|rr|rr|rr|}
			\hline
			\multirow{2}{*}{graph} & \multicolumn{2}{c|}{\# vertices} & \multicolumn{2}{c|}{\# edges} & \multicolumn{2}{c|}{time, s.}\\ 
             & single & hier. & single & hier. & single & hier. \\
             \hline
			BAV &  23153 & 21711 & 296756 & 117208 & 84.90 & 2.80 \\
			GER &  120959 & 113763 & 1447934 & 588383 & 521.75 & 14.79  \\
            NY &  51152 & 46651 & 101619 & 510550 & 61.76 & 2.89  \\
			BAY &  49385 & 42814 & 812292 & 357292 & 45.65 & 2.63  \\
            COL &  64082 & 52136 & 953940 & 396702 & 72.09 & 3.21  \\
            FLA &  168914 & 145873 & 2702892 & 1187054 & 177.65 & 9.25  \\
            NE &  241705 & 218842 & 3835060 & 1897282 & 240.36 & 15.98  \\
			\hline
		\end{tabular}
	\end{center}
\end{table}

Rather unsurprisingly, the larger the size of the graph, the more time cover construction takes, and the higher is the size difference between cover sizes. This is evident from measurements on the GER, FLA, and NE graphs. A more important observation, however, is that construction of a new level tends to take less time than that of the previous one. This is explained by the decreasing cover sizes, which means the construction algorithm has to process less and less data. The number of cover vertices falls sharply in the first several layers, after which the compression rate slows down. A more interesting picture is presented by the measurements of cover edges. As one can see, their number falls significantly in the first several iterations, but tends to grow somewhat in higher levels. The reason for that is that with the constantly decreasing number of vertices, the algorithm has to create more and more cover edges to maintain full path information inside the covers. This phenomenon can be called "oversaturation". For the majority of the tested graphs, the oversaturation threshold lies at levels 6 to 8, while for NE it starts at level 5. This observation, however, is based only on cover sizes, and one must factor in query performance on the hierarchical covers in order to make the final decision on the appropriate number of levels.

As it was stated above, one of the important features of hierarchical kPCs is the speed of their construction. The problem of constructing optimal kPC is proven to be APX-hard \cite{kPC_complexity}, heuristics methods providing good-quality solutions are used in practice. One such method is designed by Funke et.al. \cite{kPC_pruning} and used in t-discarding kPC MLS \cite{t-kPC-MLS}. It uses DFS to build a single-layer cover with a set $k$ value, and removes unnecessary cover edges using domination pruning. Information on construction times and sizes using this method is provided in \cite{t-kPC-MLS}. We used this information to compile a comparison table \ref{tab:kpc_comp}. As can be seen, the hierarchical approach outperforms the single-layer one in all aspects in our experiments. The difference in sizes of covers is, evidently, substantial. We attribute it mainly to the LR-deg strategy of cover vertex selection. In \cite{t-kPC-MLS}, an arbitrary processing order was used. The difference in construction time, on the other hand, is dramatic, and can mainly be attributed to the construction procedures. Due to 2-Path-Cover vertex and edge selection strategies being so simple and therefore extremely fast, the difference in processing times accumulates rapidly and exceeds 20 times. Moreover, constructing a single layer kPC for $k=64$ or even higher would take hours, if not days. This is because, as stated in \cite{kPC_pruning}, the processing time for kPC construction grows exponentially with the value of $k$. 

Naturally, a partial downside of hierarchical kPC is the necessity to maintain all cover layers in memory, which gives it a several times larger footprint than the single-layer one. Nevertheless, memory requirements for a large-scale multicriteria shortest path search query are incomparably higher due to the vast number of labels to be processed and maintained. Therefore, the hierarchical approach is more preferable due to its comparatively negligible construction time, making it usable for dynamic or regularly changing criteria.

\subsection{Hierarchical MLS experiments}
We tested the efficiency of Hierarchical MLS compared to classic MLS and analyzed the performance of it on every top level. The experiments were conducted on all 7 graphs listed above with 100 random source-destination pairs for each. Our measurements do not include construction time neither for hierarchical covers nor for the single-layer one. The results of experiments in terms of query times and number of produced labels are presented in Tables \ref{tab:query_times} to \ref{tab:maxquery_sizes}. Tables \ref{tab:query_times} and \ref{tab:maxquery_times} provide mean and maximum observed query times in seconds respectively, while Tables \ref{tab:query_sizes} and \ref{tab:maxquery_sizes} give the same presentation for memory requirements in millions of generated labels.

\begin{table*}[t]
	\caption{Mean query times in seconds for one-to-one MLS and Hierarchical MLS. The column names provide the top cover level, 0 being standard MLS.}
    \label{tab:query_times}
	\begin{center}
		\label{sizes}
		\begin{tabular}{|c|rrrrrrrrrrr|}
			\hline
			graph & 0 & 1 & 2& 3 & 4 & 5 & 6 & 7 & 8 & 9 & 10 \\ 
             \hline
             BAV & 32.84 & 13.71 & 7.93 & 4.95 & 3.31 & 2.51 & 2.01 & \textbf{1.80} & 1.84 & 1.85 & 1.94 \\
            GER & 909.17 & 228.15 & 154.31 & 100.10 & 69.66 & 50.07 & 43.10 & 39.78 & \textbf{38.31} & 38.70 & 40.49 \\
             NY & 102.54 & 31.87 & 24.70 & 21.44 & 18.75 & 19.28 & 19.12 & \textbf{18.51} & 19.66 & 24.73 & 25.95 \\
            BAY & 108.22 & 27.11 & 17.47 & 15.75 & 11.61 & 10.24 & 9.76 & \textbf{9.36} & 10.16 & 12.52 & 12.15 \\
            COL & 659.17 & 114.01 & 74.94 & 49.85 & 39.56 & 36.40 & 39.61 & 37.74 & \textbf{32.99} & 33.50 & 33.04 \\
            FLA & 6797.04 & 1237.69 & 858.82 & 685.50 & 572.00 & 497.65 & 456.43 & 437.91 & 404.19 & \textbf{386.64} & 395.63 \\
            NE & 6189.53 & 1346.42 & 1004.03 & 807.09 & 696.51 & 709.71 & 662.03 & 675.53 & \textbf{648.33} & 654.57 & 654.90 \\
			\hline
		\end{tabular}
	\end{center}
\end{table*}

\begin{table*}[t]
	\caption{Maximum query times in seconds for one-to-one MLS and Hierarchical MLS. The column names provide the top cover level, 0 being standard MLS.}
    \label{tab:maxquery_times}
	\begin{center}
		\label{sizes}
		\begin{tabular}{|c|rrrrrrrrrrr|}
			\hline
			graph & 0 & 1 & 2& 3 & 4 & 5 & 6 & 7 & 8 & 9 & 10 \\ 
             \hline
             BAV & 493.27 & 219.18 & 140.50 & 88.84 & 59.92 & 39.55 & 29.99 & 26.15 & 26.23 & \textbf{23.23} & 23.32 \\
            GER & 6905.68 & 831.90 & 510.08 & 331.52 & 227.17 & 167.25 & 141.40 & 130.90 & \textbf{128.13} & 134.17 & 144.02 \\
             NY & 1210.87 & 263.64 & 188.85 & 154.56 & 133.36 & 128.22 & 114.75 & 108.95 & 101.75 & \textbf{92.45} & 97.17 \\
            BAY & 1875.32 & 218.74 & 139.71 & 100.81 & 76.98 & 66.73 & 60.44 & 56.54 & \textbf{56.14} & 259.98 & 240.80 \\
            COL & 5837.71 & 659.82 & 441.01 & 323.03 & 261.89 & 221.24 & 175.47 & 163.06 & 158.02 & 151.84 & \textbf{146.34} \\
            FLA & 42733.20 & 5992.17 & 4280.66 & 3273.00 & 2772.02 & 2383.88 & 2141.79 & 2058.54 & 1976.79 & 1911.17 & \textbf{1825.51} \\
            NE & 16865.90 & 3091.01 & 2404.66 & 2032.30 & 1804.82 & 1771.07 & 1825.74 & 1827.82 & \textbf{1649.84} & 1736.87 & 1674.68 \\
			\hline
		\end{tabular}
	\end{center}
\end{table*}

\begin{table*}[t]
	\caption{Mean query sizes in millions of labels for one-to-one MLS and Hierarchical MLS. The column names provide the top cover level, 0 being standard MLS.}
    \label{tab:query_sizes}
	\begin{center}
		\label{sizes}
		\begin{tabular}{|c|rrrrrrrrrrr|}
			\hline
			graph & 0 & 1 & 2& 3 & 4 & 5 & 6 & 7 & 8 & 9 & 10 \\ 
             \hline
             BAV & 3.77 & 2.04 & 1.15 & 0.67 & 0.42 & 0.28 & 0.20 & 0.16 & 0.13 & 0.11 & \textbf{0.09} \\
            GER & 70.74 & 38.38 & 21.66 & 12.83 & 8.05 & 5.41 & 3.89 & 2.98 & 2.40 & 2.02 & \textbf{1.75} \\
             NY & 7.52 & 4.16 & 2.70 & 1.94 & 1.50 & 1.22 & 1.04 & 0.91 & 0.81 & 0.73 & \textbf{0.67} \\
            BAY & 7.40 & 3.81 & 2.29 & 1.50 & 1.07 & 0.81 & 0.65 & 0.54 & 0.46 & 0.41 & \textbf{0.36} \\
            COL & 27.33 & 14.02 & 8.19 & 5.22 & 3.60 & 2.67 & 2.09 & 1.70 & 1.44 & 1.25 & \textbf{1.10} \\
            FLA & 143.29 & 73.68 & 43.86 & 28.45 & 19.81 & 14.67 & 11.43 & 9.26 & 7.75 & 6.64 & \textbf{5.81} \\
            NE & 171.69 & 91.26 & 56.54 & 38.54 & 28.42 & 22.28 & 18.29 & 15.55 & 13.56 & 12.06 & \textbf{10.90} \\
			\hline
		\end{tabular}
	\end{center}
\end{table*}

\begin{table*}[t]
	\caption{Maximum query sizes in millions of labels for one-to-one MLS and Hierarchical MLS. The column names provide the top cover level, 0 being standard MLS.}
    \label{tab:maxquery_sizes}
	\begin{center}
		\label{sizes}
		\begin{tabular}{|c|rrrrrrrrrrr|}
			\hline
			graph & 0 & 1 & 2& 3 & 4 & 5 & 6 & 7 & 8 & 9 & 10 \\ 
             \hline
             BAV & 39.45 & 21.39 & 11.91 & 6.91 & 4.23 & 2.77 & 1.95 & 1.47 & 1.17 & 0.98 & \textbf{0.84} \\
            GER & 274.26 & 148.35 & 83.46 & 49.29 & 30.90 & 20.79 & 15.05 & 11.57 & 9.38 & 7.91 & \textbf{6.90} \\
            NY & 64.93 & 34.89 & 22.01 & 15.31 & 11.51 & 9.18 & 7.64 & 6.55 & 5.74 & 5.13 & \textbf{4.65} \\
            BAY & 65.16 & 33.33 & 19.86 & 12.95 & 9.13 & 6.86 & 5.42 & 4.47 & 3.81 & 3.32 & \textbf{2.94} \\
            COL & 180.09 & 93.07 & 55.22 & 35.75 & 25.01 & 18.69 & 14.69 & 12.06 & 10.22 & 8.90 & \textbf{7.89} \\
            FLA & 637.42 & 332.03 & 200.71 & 132.22 & 93.18 & 69.57 & 54.52 & 44.29 & 37.07 & 31.80 & \textbf{27.82} \\
            NE & 330.14 & 175.65 & 108.37 & 73.58 & 54.05 & 42.27 & 34.63 & 29.40 & 25.63 & 22.78 & \textbf{20.57} \\
			\hline
		\end{tabular}
	\end{center}
\end{table*}

Analyzing this data leads us to the following conclusions. The number of created labels steadily decreases with the increase in the number of cover levels. This is caused by the steadily decreasing number of vertices in the covers. For a given source-destination pair, every vertex has a Pareto set of a specific size for it. However, the overwhelming majority of the vertices that do not belong to the top cover are skipped during the query, and thus no labels are generated for them. Increasing the number of cover levels to a certain degree is thus a reliable way to decrease memory requirements of shortest path search queries. Unfortunately, the situation is not as straightforward with operation time. Although any hierarchically enabled MLS is faster than the one conducted on the original graph, the best running times are achieved by queries at levels 7 to 9, with level 10 providing slightly better performance in some extreme cases. These results tend to correspond to the numbers of edges in cover levels, leading us to the conclusion that the number of edges has a greater effect on running time than the number of vertices. This tendency, however, does not always hold (as is evident from the test results on NE graph) and should be followed cautiously. Nevertheless, setting a corresponding level limit presents itself to be a convenient way to achieve a desired balance between memory consumption and processing time. 

At best levels, Hierarchical MLS provides average speedups ranging from 5 times for NY to 23 times for GER. Average memory requirements, on the other hand, were reduced by 11 times for NY to 42 times for GER. From these observations, we conclude level 8, corresponding to 256-Path-Cover, to be the safest bet if one is not willing to perform a preliminary empirical analysis on their problem instance. Interestingly, one-to-one $t$-discarding kPC-MLS in \cite{t-kPC-MLS} was reported to provide speedups of 2.96, 5.51, and 8.67 for its best-performing graphs NY, BAY, and COL respectively. According to our measurements, Hierarchical MLS can provide speedups up to 5.54, 11.56, and 19.98 on the same graphs. The superiority of Hierarchical MLS stems from significantly better compression rates for hierarchical covers and the hierarchical approach being able to provide covers for $k$ values unachievable for the alternative method.

\section{Conclusion}
This paper presents a new multicriteria shortest path search algorithm. The novelty of Hierarchical MLS lies in its operation on a structure of nested $k$-Path-Covers. This feature makes it not only applicable to dynamic and apriori-unknown criteria, but allows it to be conveniently and reliably adjusted to achieve the desired time/memory trade-off. The benefit provided by our algorithm does not rely on heuristic metrics or pruning procedures, making it a robust option for optimal multicriteria shortest path search. Furthermore, it is able to incorporate additional speedup techniques with minimal modification required. We test the algorithm on several graphs of varying sizes and compare it to its closest alternatives. The results of our experiments confirm Hierarchical MLS to be superior in terms of speedup, memory efficiency, and flexibility.

\bibliographystyle{IEEEtran}
\bibliography{bibliography}

\end{document}